\documentclass[superscriptaddress,twocolumn,showpacs,preprintnumbers,amsmath,amssymb,longbibliography,pra]{revtex4-1}
\usepackage{graphicx}
\usepackage{dcolumn}
\usepackage{bm}
\usepackage[scriptsize]{subfigure}
\usepackage{float}
\usepackage{amsmath}
\usepackage{lipsum}
\usepackage{mathtools}

\renewcommand{\vec}[1]{\boldsymbol{#1}}

\begin{document}


\title{Doubly-differential cross section calculations for $K$-shell vacancy production in lithium by fast O$^{8+}$ ion impact}

\author{M. D. \'Spiewanowski}
\email[]{macieks@yorku.ca}
\affiliation{Department of Physics and Astronomy, York University, Toronto, Ontario, Canada M3J 1P3}
\author{L. Guly\'as}
\affiliation{Institute for Nuclear Research, Hungarian Academy of Sciences (ATOMKI), P.O. Box 51, H-4001 Debrecen, Hungary}
\author{M. Horbatsch}
\affiliation{Department of Physics and Astronomy, York University, Toronto, Ontario, Canada M3J 1P3}
\author{T. Kirchner}
\email[]{tomk@yorku.ca}
\affiliation{Department of Physics and Astronomy, York University, Toronto, Ontario, Canada M3J 1P3}
\date{\today}

\begin{abstract}
Inner-shell vacancy production for the O$^{8+}$-Li collision system at 1.5~MeV/amu is studied theoretically. 
The theory combines single-electron amplitudes for each electron in the system to extract multielectron information about the collision process.
Doubly-differential cross sections obtained in this way are then compared with the recent experimental data by LaForge et al. [J. Phys. B 46, 031001 (2013)] 
yielding good resemblance, especially for low outgoing electron energy. A careful analysis of the processes that contribute to inner-shell vacancy production shows that 
the improvement of the results as compared to single-active-electron calculations can be attributed to the leading role of two-electron
excitation-ionization processes.
\end{abstract}

\pacs{34.50.Fa}

\maketitle
\section{Introduction}\label{intro}
Single, as well as multiple ionization processes of atoms by fast bare ion impact have been of great interest for decades~\cite{Schulz2006} 
since they provide basic information about mechanisms for few-body breakup processes. 
Experimentally, the development of cold-target recoil-ion momentum spectroscopy (COLTRIMS)~\cite{UllrichRepProgPhys2003} allows to study the collision processes in great detail.
Yet, due to the limitations posed by the supersonic gas jet target technique, the targets of interest are typically limited to rare gas atoms and molecules~\cite{SchulzNature2003}.

In recent experiments, laser cooling in a magneto-optical trap was combined with a reaction microscope 
(MOTReMi)~\cite{FischerPRL2012,HubeleRSI2015} to study single and multiple ionization of alkali metal atoms by ion impact. 
With this new approach the first kinematically complete experiment for single ionization of lithium was achieved and experiments were performed for the H$^{+}$-Li collision system at 6~MeV/amu~\cite{LaForgeJPB2013}, the 
O$^{8+}$-Li collision system at 1.5~MeV/amu~\cite{LaForgeJPB2013,HubelePRL2013}, and the 
Li$^{2+}$-Li collision system at 2.29~MeV/amu~\cite{SpiewanowskiJPCF2015}. This 
triggered theoretical efforts for the interpretation of the experimental data~\cite{CiappinaPRA2013,WaltersPRA2014,KirchnerPRA2014,GulyasPRA2014,SpiewanowskiJPCF2015}. 

The 'hydrogen-like' structure of the lithium atom with its large energy spacing between the $K$- and $L$-shells allows for efficient laser cooling in a MOT and, hence, for
accurate studies of single-ionization processes. At the level of the \textit{singly} differential cross section (SDCS) outer-shell ionization has proven to be nearly 
a single-electron process with a marginal influence of the $K$-shell electrons~\cite{KirchnerPRA2014}. Yet, \textit{doubly} differential cross sections (DDCSs) for outer-shell ionization
show that the inner-shell electrons do play a role in the process, e.g., via core polarization by the projectile~\cite{GulyasPRA2014}. 
In $K$-shell vacancy production, on the other hand, the effects of the presence of all electrons are important already at the level of the SDCS~\cite{KirchnerPRA2014}.

Here, we study cross sections differential in energy of the outgoing electron and in the transverse momentum transfer by applying a novel strategy of combining single-electron amplitudes.
In this way we incorporate the multielectron nature of the collision process into the theoretical model. We apply the two-center basis generator method (TC-BGM)~\cite{ZapukhlyakJPB2005} 
to calculate amplitudes for electron transitions within the target atom, and combine them with continuum distorted wave eikonal initial state (CDW-EIS)~\cite{GulyasJPB1995} 
single-electron ionization amplitudes. 
The combined amplitudes are then amended with a phase factor that accounts for the nucleus-nucleus interaction, and are subsequently 
Fourier transformed in order to obtain the DDCSs differential in the outgoing-electron energy and in the transverse momentum transfer.
When compared with experimental data our results show improvements relative to previous attempts and lead to a better understanding 
of the underlying processes that take place during inner-shell vacancy production for this collision system.

The paper is organized as follows. The novel methodology to extract multielectron information from single-particle amplitudes is given in Sec.~\ref{Sec:theory}.
In Sec.~\ref{Sec:results} we present DDCSs for the O$^{8+}$-Li system at 1.5~MeV/amu together with a careful analysis of the contributions that constitute the overall DDCS.
We draw conclusions in Sec.~\ref{Sec:conclusions}. We use atomic units ($\hbar=m_e=e=4 \pi \epsilon_0 = 1$) throughout the paper unless stated otherwise.

\section{Methodology}\label{Sec:theory}
We use the framework of the semiclassical approximation, in which the projectile follows a straight-line trajectory $\vec{R}=\vec{\rho}+\vec{v}t$ with impact parameter vector $\vec{\rho}$
and constant velocity $\vec{v}$.
The cross section differential in the outgoing-electron energy $E_e$ and in the transverse momentum transfer $\eta$ is given as 
\begin{equation}\label{DDCS}
 \frac{\mathrm{d}\sigma}{\mathrm{d}E_e\mathrm{d}\eta} = k_e\eta\int\limits_{-1}^1d(\cos\theta_e)\int\limits_0^{2\pi}d\phi_e\int\limits_0^{2\pi}d\phi_\eta|R_{i\vec{k_e}}(\vec{\eta})|^2,
\end{equation}
with the transition matrix element $R_{i\vec{k_e}}(\vec{\eta})$ for the transition from the initial state to a state that includes an unbound electron with momentum $\vec{k_e}$
\begin{equation}\label{2DFourier}
 R_{i\vec{k_e}}(\vec{\eta}) = \frac{1}{2\pi}\int d\vec{\rho} e^{i\vec{\eta}\cdot\vec{\rho}}a_{i\vec{k_e}}(\vec{\rho}),
\end{equation}
and
\begin{equation}\label{aik}
a_{i\vec{k_e}}(\vec{\rho})=e^{i\delta(\rho)}A_{i\vec{k_e}}(\vec{\rho}).
\end{equation}
Here $A_{i\vec{k_e}}(\vec{\rho})$ is the impact-parameter-dependent multielectron amplitude calculated without the influence of the internuclear interaction with details given below. 
The phase accumulated during the collision process due to the projectile interaction with the target nucleus, 
$\delta(\rho)$, is given by (see, e.g., Ref.\cite{ZapukhlyakJPB2008})
\begin{equation}\label{delta}
\delta(\rho)=-\int\limits_{-\infty}^\infty dt V_\mathrm{NN}(R(t)),
\end{equation}
with the nucleus-nucleus (NN) potential
\begin{equation}\label{vnn}
V_\mathrm{NN}(R)=\frac{Z_PZ_T}{R}.
\end{equation}

Within the independent-electron (IEL) model the full three-electron Hamiltonian of the collision problem is replaced by a sum of single-particle operators 
\begin{equation}\label{IELHam}
H_e(t)=\sum\limits_{i=1}^3 \left(-\frac{1}{2}\Delta_i(t)+V\left(r_t^i\right)-\frac{Z_p}{r_p^i}\right).
\end{equation}
The last two terms in Eq.~\eqref{IELHam} describe the interaction of the $i$th electron with the target and the projectile, respectively; $r_t^i$ and $r_p^i$ are the distances between
the $i$th electron and the target and the projectile, respectively, and $Z_p$ is the projectile charge. 

The TDSE with the Hamiltonian given by Eq.~\eqref{IELHam} separates into a set of single-particle equations whose solutions are used to calculate electronic transition amplitudes 
for the three-electron system in a completely specified final state. In order to do so we assemble the single-particle solutions to form a Slater determinant 
($|\psi_{1s\uparrow}\psi_{1s\downarrow}\psi_{2s\uparrow}(\vec{\rho},t)\rangle$), where the subscripts indicate the initial single-electron states. 
We project it at $t=t_f$ onto the three-electron final-state Slater determinant ($|\alpha_{1\uparrow}\alpha_{2\downarrow}\alpha_{3\uparrow}\rangle$), where $\alpha_{i}\ (i=1,2,3)$ 
are the final states of interest~\cite{LuddeJPB1985}
\begin{equation}\label{A}
A_{\alpha_{1\uparrow}\alpha_{2\downarrow}\alpha_{3\uparrow}}(\vec{\rho})=\langle\alpha_{1\uparrow}\alpha_{2\downarrow}\alpha_{3\uparrow}|\psi_{1s\uparrow}\psi_{1s\downarrow}\psi_{2s\uparrow}(\vec{\rho},t_f)\rangle.
\end{equation}

The overall DDCS for the process of interest to this work, namely, vacancy production in the inner shell (with the detached electron having a well-defined momentum $\vec{k_e}$)
is obtained by plugging Eq.~\eqref{A} into Eq.~\eqref{aik} and working through Eqs.~\eqref{2DFourier} and \eqref{DDCS} for every possible final state. The result can be expressed as follows:
\begin{equation}\label{Pvacke1}
\frac{\mathrm{d}\sigma^\mathrm{vac}}{\mathrm{d}E_e\mathrm{d}\eta}=\sum\limits_{f\neq 1s}\left(\frac{\mathrm{d}\sigma_{1s\uparrow{f\downarrow}\vec{k_e}\uparrow}}{\mathrm{d}E_e\mathrm{d}\eta} 
+ \frac{\mathrm{d}\sigma_{f\uparrow1s\downarrow\vec{k_e}\uparrow}}{\mathrm{d}E_e\mathrm{d}\eta} + \frac{\mathrm{d}\sigma_{1s\uparrow\vec{k_e}\downarrow{f}\uparrow}}{\mathrm{d}E_e\mathrm{d}\eta}\right).
\end{equation}
Here $\frac{\mathrm{d}\sigma_{\alpha_1\alpha_2\alpha_3}}{\mathrm{d}E_e\mathrm{d}\eta}$ is the DDCS for the process with the final state indicated by $\alpha_{i}\ (i=1,2,3)$ with the proper spin projections, 
and the sum runs over the target bound-state manifold. After some simplifications that are outlined in the Appendix we arrive at 
\begin{eqnarray}\label{Pvacke2}
\frac{\mathrm{d}\sigma^\mathrm{vac}}{\mathrm{d}E_e\mathrm{d}\eta}&=&\frac{\mathrm{d}\sigma^\mathrm{excl}}{\mathrm{d}E_e\mathrm{d}\eta}+\sum\limits_{\mathclap{f\neq 1s,2s}}\frac{\mathrm{d}\sigma^\mathrm{EI1}_{f}}{\mathrm{d}E_e\mathrm{d}\eta}
+\sum\limits_{f\neq 1s}\frac{\mathrm{d}\sigma^\mathrm{EI2}_{f}}{\mathrm{d}E_e\mathrm{d}\eta}\nonumber\\
&&+\sum\limits_{f\neq 1s}\frac{\mathrm{d}\sigma^\mathrm{ex}_{f}}{\mathrm{d}E_e\mathrm{d}\eta} + \Delta_{E_e,\eta},
\end{eqnarray}
where we can distinguish the following processes and amplitudes associated with them:

(i) Exclusive ionization, where one $1s$ electron is removed, whereas the two electrons that are left behind remain unaffected
\begin{equation}\label{excl}
A^\mathrm{excl}=\sqrt{2}A_{1s\rightarrow 1s}A_{1s\rightarrow \vec{k_e}}A_{2s\rightarrow 2s};
\end{equation}

(ii) Excitation-ionization (EI1), with one-electron removal from the inner shell, one-electron excitation from the outer shell, and the residual inner-shell electron remaining unaffected
\begin{equation}\label{EI1}
A_{f}^\mathrm{EI1}=\sqrt{2}A_{1s\rightarrow 1s}A_{1s\rightarrow \vec{k_e}}A_{2s\rightarrow f};
\end{equation}

(iii) Excitation-ionization (EI2), with one-electron removal from the outer shell, one-electron excitation from the inner shell, and the residual inner-shell electron remaining unaffected
\begin{equation}\label{EI2}
A_{f}^\mathrm{EI2}=\sqrt{2}A_{1s\rightarrow 1s}A_{1s\rightarrow f}A_{2s\rightarrow \vec{k_e}}.
\end{equation}
The last two terms on the right-hand side of Eq.~\eqref{Pvacke2}, correspond to exchange processes, $\frac{\mathrm{d}\sigma^\mathrm{ex}_{f}}{\mathrm{d}E_e\mathrm{d}\eta}$, 
and an antisymmetry correction term $\Delta_{E_e,\eta}$, that contains all the cross terms. The origin of these terms is explained in the Appendix. 
They turn out to be negligibly small and, henceforth, will be omitted~\cite{KirchnerPRA2014}.

(iv) Shake off. In addition to the above, we introduce this correlated two-electron process that has proven to be of importance for the creation of low-energy electrons during inner-shell 
vacancy production~\cite{KirchnerPRA2014}. One $1s$ electron is excited while the $2s$ electron is shaken off ending up in the continuum, 
whereas the residual inner-shell electron remains unaffected
\begin{equation}\label{shake}
A_{f}^\mathrm{shake}=\sqrt{2}A_{1s\rightarrow 1s}A_{1s\rightarrow f}A_{2s}^\mathrm{overlap}.
\end{equation}
Following~\cite{KirchnerPRA2014} we employed the independent event (IEV) model for the EI2 process where, in our calculation, ionization takes place from the valence shell of the target in the ground-state
configuration, and the elastic scattering and excitation processes take place after the inner electrons rearrange to the Li$^+$ ground-state configuration. 

The amplitudes on the right-hand side of Eqs.~\eqref{excl}-\eqref{shake} are single-particle transition amplitudes and $\sqrt{2}$ 
accounts for the indistinguishability of the electrons in the $K$-shell (see Appendix). In Eq.~\eqref{shake} $A_{2s}^\mathrm{overlap}$ is the overlap between the outer-shell electron orbital 
and a continuum state of a modified Hamiltonian with the proper outgoing-electron energy~\cite{KirchnerPRA2014}.

Transition amplitudes for electron excitations for processes (i)-(iv) are calculated using the two-center basis generator method (TC-BGM)~\cite{ZapukhlyakJPB2005}. 
The interactions within the target atom [see Eq.~\eqref{IELHam}] have been approximated by the exchange-only optimized potential method (OPM) of density functional theory~\cite{EngelPRA1993}. 
Single-particle TDSEs are solved using a basis that consists of $1s-4f$ target states, and $1s-4f$ projectile states together 
with 71 BGM pseudostates~\cite{ZapukhlyakJPB2005}. Electron capture has been shown to be very weak in this collision system~\cite{KirchnerPRA2014}, but the projectile states are included 
nevertheless to ensure consistency with the calculation of Ref.~\cite{KirchnerPRA2014}.
All basis states are endowed with electron translation factors to ensure Galilean invariance.
Amplitudes for ionization, on the other hand are obtained from CDW-EIS, due to the difficulties posed by the extraction of these amplitudes from TC-BGM.
We note that the ionization probabilities calculated from both methods are in very good agreement (see Fig. 1 in Ref.~\cite{KirchnerPRA2014}).

\begin{figure}[t]
\centering
\includegraphics[width=0.51\textwidth]{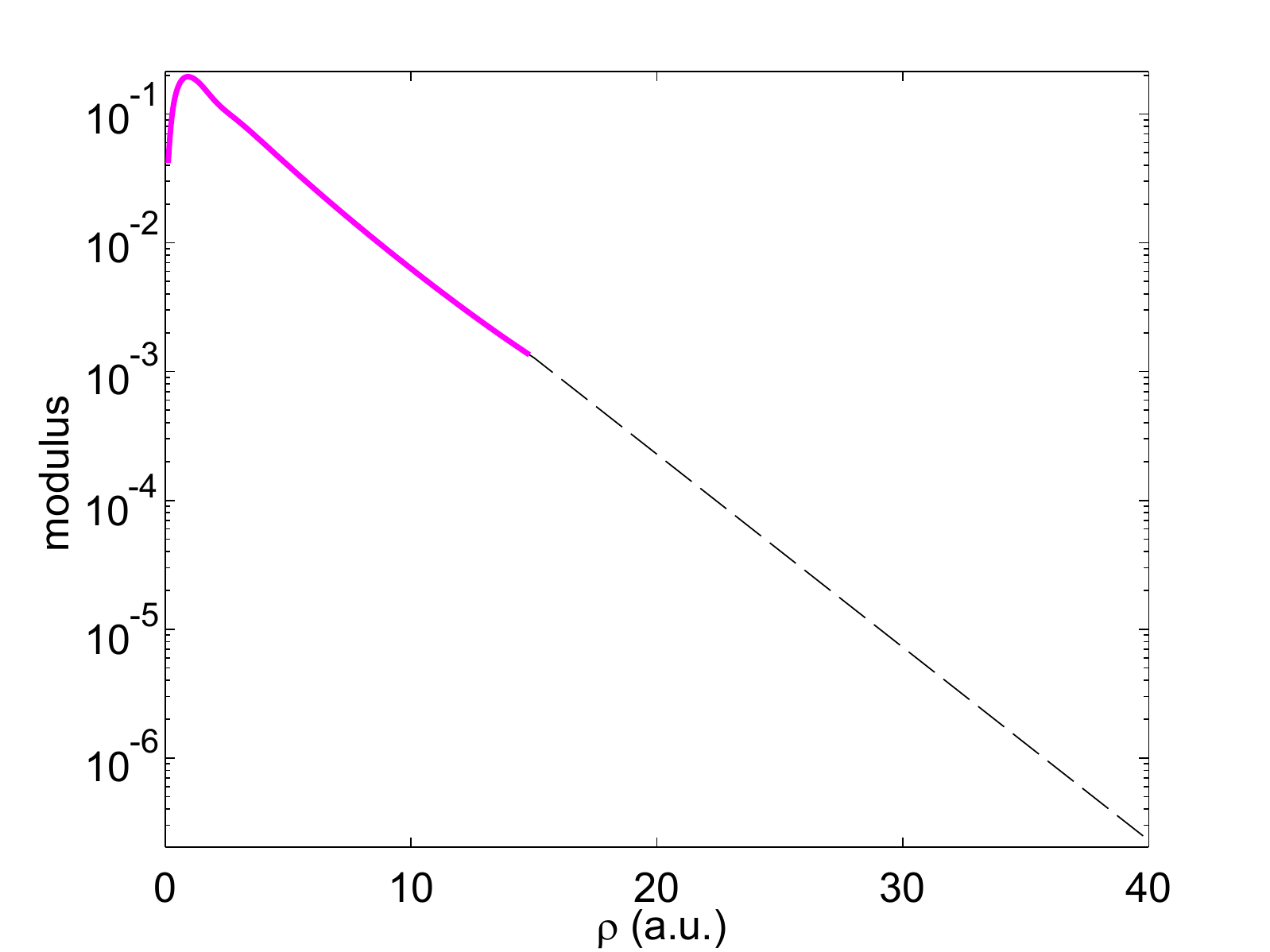}
\caption{(Color online) Modulus of the 1$s\rightarrow2p_0$ single-particle transition amplitude as a function of impact parameter. Solid line (magenta) - TC-BGM; dashed line (black) - fit.}
\label{Abs}
\end{figure}

\begin{figure}[t]
\centering
\includegraphics[width=0.51\textwidth]{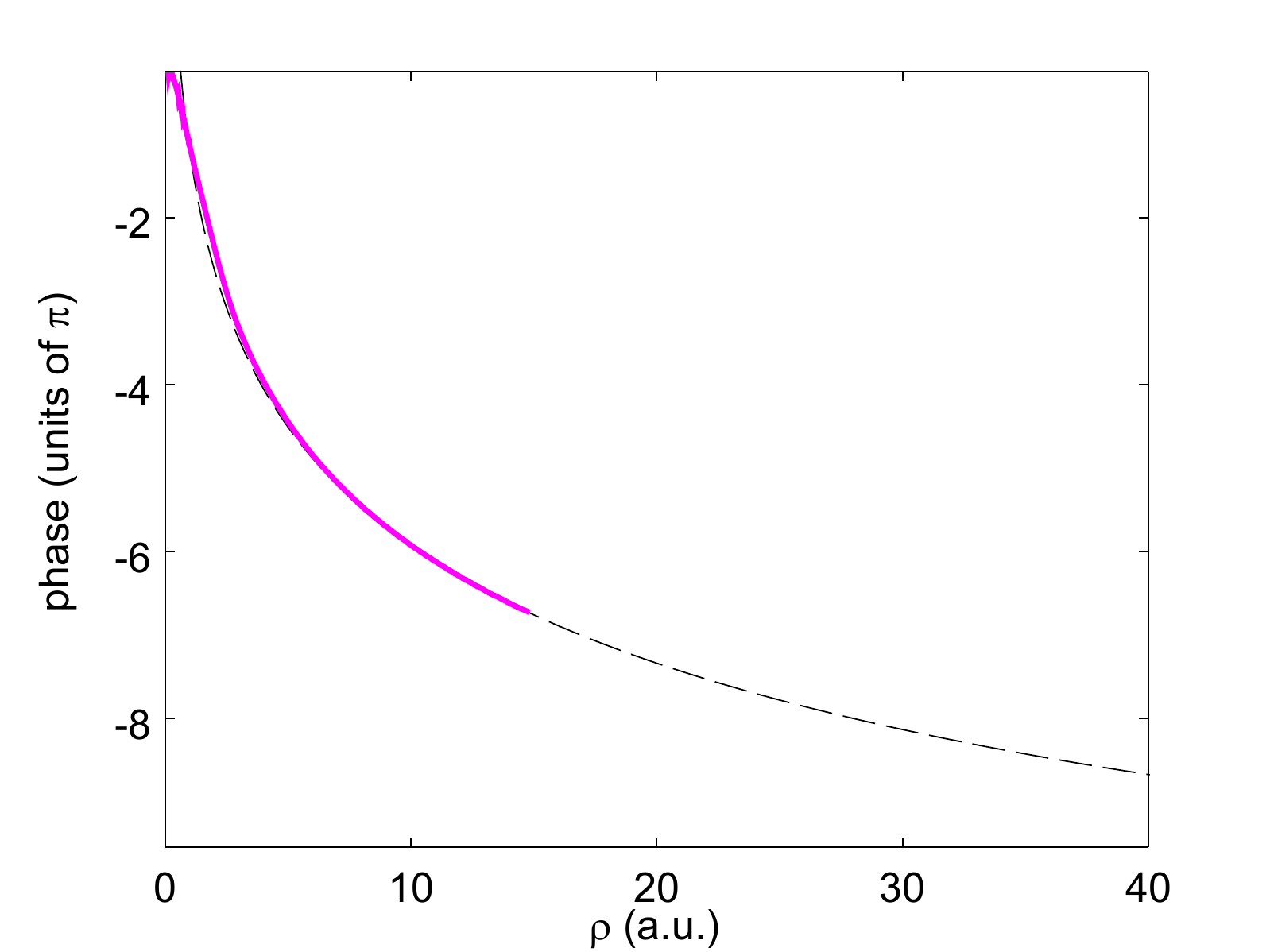}
\caption{(Color online)  Phase of the 1$s\rightarrow2p_0$ single-particle transition amplitude as a function of impact parameter. Solid line (magenta) - TC-BGM; dashed line (black) - fit.}
\label{phase}
\end{figure}

For the Fourier transform of Eq.~\eqref{2DFourier} to be fully converged one needs the three-electron amplitudes~\eqref{excl}-\eqref{shake} for impact parameters 
up to a few hundred atomic units. Unlike the CDW-EIS, the TC-BGM calculations give reliable results only for a limited range of impact parameters. 
It is, therefore, necessary to extrapolate the excitation amplitudes from TC-BGM calculations, both their moduli and phases, to avoid numerical issues within the Fourier transform procedure. 
In Figs.~\ref{Abs} and ~\ref{phase} we depict an example of the absolute value and phase, respectively, of a transition amplitude from TC-BGM 
together with a fit for a range of impact parameters. 
For the purpose of this work such fits were carried out for each excitation channel.

The sums in Eq.~\eqref{Pvacke2} run over the target bound-state
manifold. 
Due to the restricted basis in the TC-BGM calculations we truncate these sums at $n=4$, where $n$ is the principal quantum number. Even after the truncation 
the procedure is still very demanding numerically with the main challenge posed by the 2D Fourier transform of three-electron amplitudes in Eq.~\eqref{2DFourier} that has to be calculated for each channel.

We note that in most of the previous calculations for the lithium MOTReMI experiments the single-active-electron (SAE) model was used. 
It is important to stress that in the SAE one does not account for interactions of the projectile with the nucleus
or the passive electrons in the single-particle transition amplitude. 
Therefore, one should amend it by a phase factor that incorporates a potential, $V_\mathrm{NN}$, 
that accounts for these interactions, i.e., $V_\mathrm{NN}=\frac{Z_PZ_T}{R}+V_{\mathrm{scr}}(R)+V_{\mathrm{pol}}(R)$, where the last two terms on the right-hand side are 
the screening potential of the passive electrons and a polarization potential, respectively, see, e.g.,~\cite{GulyasPRA2014}. It has been shown that one has to properly take into account 
the phase accumulated via $V_\mathrm{NN}$ for a complete understanding of the outer-shell ionization process~\cite{GulyasPRA2014}. In our IEL model it would be inconsistent 
to include screening and polarization terms in $V_\mathrm{NN}(R)$, since all electrons are active. $V_{\mathrm{scr}}(R)$ is inherently accounted for in this framework~\cite{ZapukhlyakJPB2008} 
while polarization would have to be modeled in terms of a time-dependent effective potential in the single-particle Hamiltonian.

Finally, we note that a similar approach has been employed earlier to calculate SDCSs where, however, probabilities, instead of amplitudes, for different transitions have been combined.
In that case it was not necessary to carry out the 2D Fourier transform.
The approach allowed to explain SDCSs for the O$^{8+}$-Li collision system at 1.5~MeV/amu~\cite{KirchnerPRA2014}. 
However, it did not agree well with the experimental observations for the case of Li$^{2+}$-Li collisions at 2.29~MeV/amu~\cite{SpiewanowskiJPCF2015}.

\begin{figure}[t]
\centering
\includegraphics[width=0.51\textwidth]{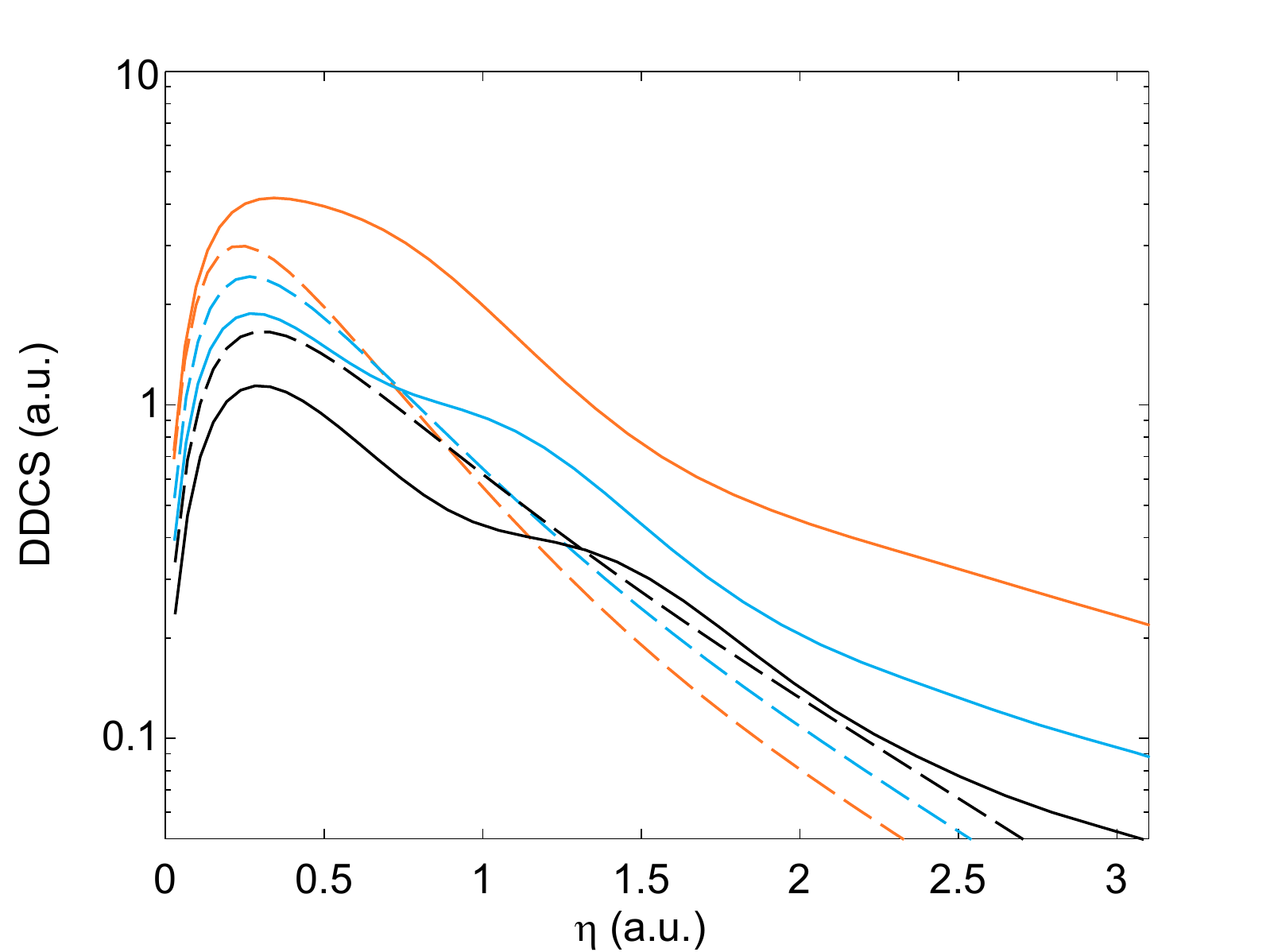}
\caption{(Color online) DDCS for the 1.5~MeV/amu O$^{8+}$-Li collision system as a function of impact parameter for three outgoing-electron energies: $E_e=2$~eV (orange), 10~eV (blue), 20~eV (black). 
Solid lines - present overall results, dashed lines - CDW-EIS in SAE approximation.}
\label{fig1a}
\end{figure}


\section{Results}\label{Sec:results}

The method just outlined gives access to calculating DDCSs differential in the outgoing-electron energy $E_e$ and in the transverse momentum transfer $\eta$. This is not possible 
when having transition probabilities as a starting point (see Ref.~\cite{KirchnerPRA2014}). 
The SDCSs can be calculated by integrating the DDCSs of Eq.~\eqref{DDCS} over the transverse momentum transfer for every $E_e$. 
To connect our work to the previous findings we cross-checked our overall SDCS results as well as results for the SDCSs for the contributing processes
with those of~\cite{KirchnerPRA2014} and concluded that the agreement between the two approaches is very good (not shown).

\begin{figure}[t]
\centering
\includegraphics[width=0.51\textwidth]{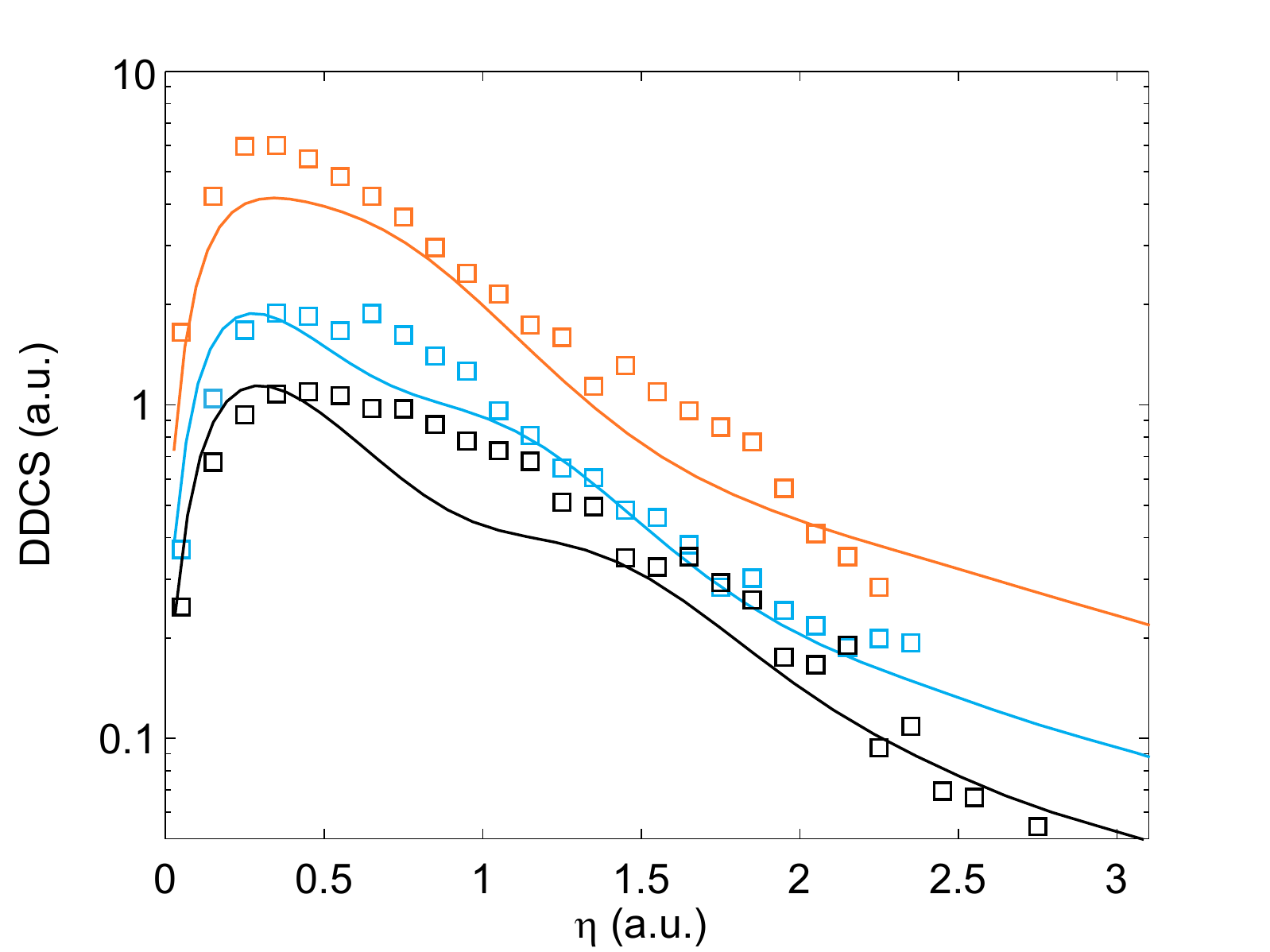}
\caption{(Color online) DDCS for the 1.5~MeV/amu O$^{8+}$-Li collision system as a function of impact parameter for three outgoing-electron energies: $E_e=2$~eV (orange), 10~eV (blue), 20~eV (black). 
Squares - experimental data from Ref.~\cite{LaForgeJPB2013}, solid lines - present overall results.}
\label{fig1b}
\end{figure}


Now we turn our attention to the results for inner-shell vacancy production obtained from the procedure outlined in the previous section. 
We compare the results of the present calculations with the SAE results
from the CDW-EIS model and with the experimental data.

The original experimental data from Ref.~\cite{LaForgeJPB2013} are relative only. In Ref.~\cite{FischerPRL2012} the SDCSs were normalized using SAE CDW-EIS for the 2s ionization channel for which the multielectron effects were found to be unimportant. 
The SDCS is obtained by integrating the experimental DDCS for the lowest $E_e=2$~eV over the transverse momentum transfer. 
It is then normalized to match the experimental value of the SDCS from Ref.~\cite{FischerPRL2012}.
This procedure fixes the normalization of all channels (in $K$-shell ionization for $E_e=2$, 10, 20~eV).

Finally, we carefully study the contributions to the overall DDCS from the processes (i)-(iv) introduced in  the previous section.

\subsection{DDCSs}\label{Sec:DDCS}

In Fig.~\ref{fig1a} we compare our overall results for electron energies $E_e=2$, 10, 20~eV (solid lines) with SAE CDW-EIS results (dashed lines). 
When comparing with Ref.~\cite{LaForgeJPB2013} we note that our own SAE CDW-EIS results agree very well in shape but we do not apply any arbitrary factors to change the spacing between the curves for all three energies. 

The SDCS differential in the energy of the outgoing electron was shown to have a very weak dependence on the electron energy in the range of interest when calculated in the SAE model~\cite{KirchnerPRA2014}. 
This trend is also seen in the case of the DDCSs, i.e., the dashed lines are very close to each other. The SAE CDW-EIS theory predicts that for low transverse momentum transfers
(which classically correspond to large impact parameters, i.e., distant collisions) it is more likely to produce low-energy electrons, and for high transverse momentum transfers one obtains more high-energy electrons. 
The orange (2~eV) and blue (10~eV) dashed curves cross at transverse momentum transfer $\eta\approx0.67$~a.u., 
and the blue (10~eV) and black (20~eV) dashed curves cross at around $\eta\approx1.10$~a.u..

After the multielectron processes have been incorporated the SDCS decreases much more rapidly with increasing outgoing-electron energy~\cite{KirchnerPRA2014}.
At the level of the DDCS this leads to the cross sections being more separated from each other, as seen in Fig.~\ref{fig1a} (solid lines). Moreover, 
the lines never cross and tend to be parallel at $\eta>2.5$~a.u.. 
Hence, slow electrons are more likely to be released than the fast ones for the whole range of $\eta$.

\begin{figure}[t]
\centering
\includegraphics[width=0.51\textwidth]{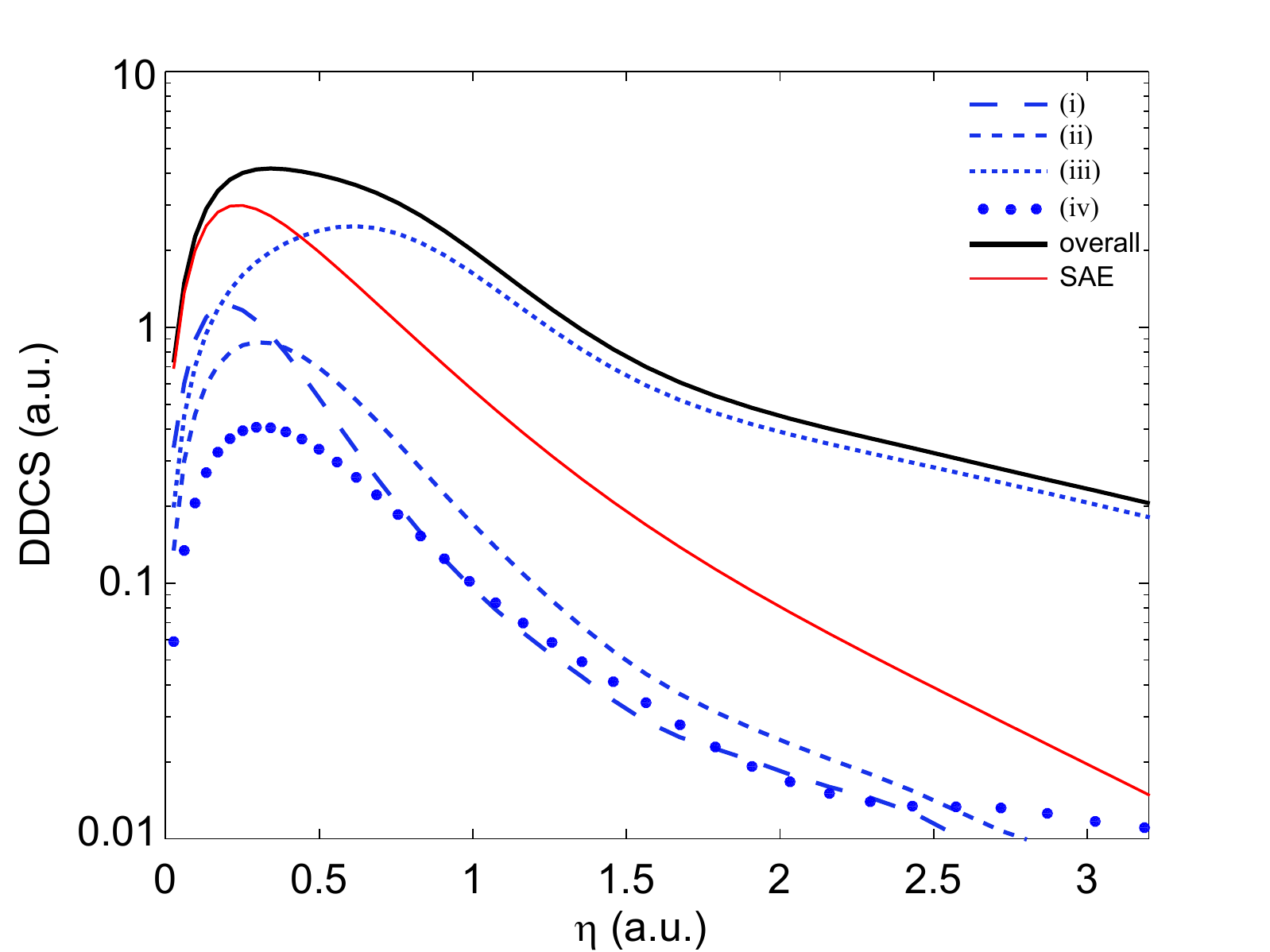}
\caption{(Color online) Partial contributions to the DDCS for processes (i)-(iv), the overall DDCS, and SAE CDW-EIS results for the O$^{8+}$-Li system at 1.5~MeV/amu collision for outgoing-electron energy $E_e= 2$~eV.}
\label{fig2}
\end{figure}

Our results are compared with the experimental data in Fig.~\ref{fig1b}.
The similarity in shape of the theoretical and experimental results for $E_e=2$~eV up to $\eta\approx2.2$~a.u. is striking. The general trend of the theoretical curve lying below the experimental points 
is attributed to the fact that only a limited number of excitation channels have been accounted for (up to $n=4$). In the case of the SDCS 
incorporation of all excitation channels in an approximate way increased the cross section by 20\%~\cite{KirchnerPRA2014}. 
This fact, we believe, explains the small difference between theory and experiment for low momentum transfers. 
An analogous approximation on the level of amplitudes is not possible without introducing arbitrary phases of excitation amplitudes causing unpredictable changes in the DDCSs.
Therefore, we refrain from extrapolating the excitation amplitudes for $n>4$.

The sudden decrease of DDCSs at around $\eta=2$~a.u. at small electron energies seen in~\cite{LaForgeJPB2013} is now deemed to be an artifact caused by the limited experimental momentum acceptance~\cite{privateFisher}.
Therefore, we have removed the three experimental points for $E_e=2$ and 10~eV for the highest transverse momentum transfer (cf. Fig.~\ref{fig1b} and Fig.~3b from~\cite{LaForgeJPB2013}). 


The agreement between theory and experiment is less satisfying for the two higher outgoing-electron energies ($E_e=10$ and 20~eV). While the general trend of the experimental results appears to be reproduced
the extended plateaus between $\eta=0.25$~a.u. to $\eta=0.75$~a.u. and $\eta=0.25$~a.u. to $\eta=1.0$~a.u. for $E_e=10$~eV and $E_e=20$~eV, respectively, are not accounted for properly. 
Instead of the aforementioned plateaus the theoretical results display narrower maxima followed by 'knees' that widen the distribution over the momentum transfer as compared to the SAE CDW-EIS results.

We elaborate more on the origin of the maxima and knees in the next subsection. For momentum transfers between the maxima and knees there is a region for which a large discrepancy between theory and experiment
is observed. Moreover, similarly to the case of the lowest outgoing-electron energy the theory overestimates the experimental data for high momentum transfers, 
especially for $E_e=10$~eV. Nevertheless, there is significant improvement as compared to the SAE CDW-EIS results allowing us to conclude that we qualitatively understand the experimental data.

\begin{figure}[t]
\centering
\includegraphics[width=0.51\textwidth]{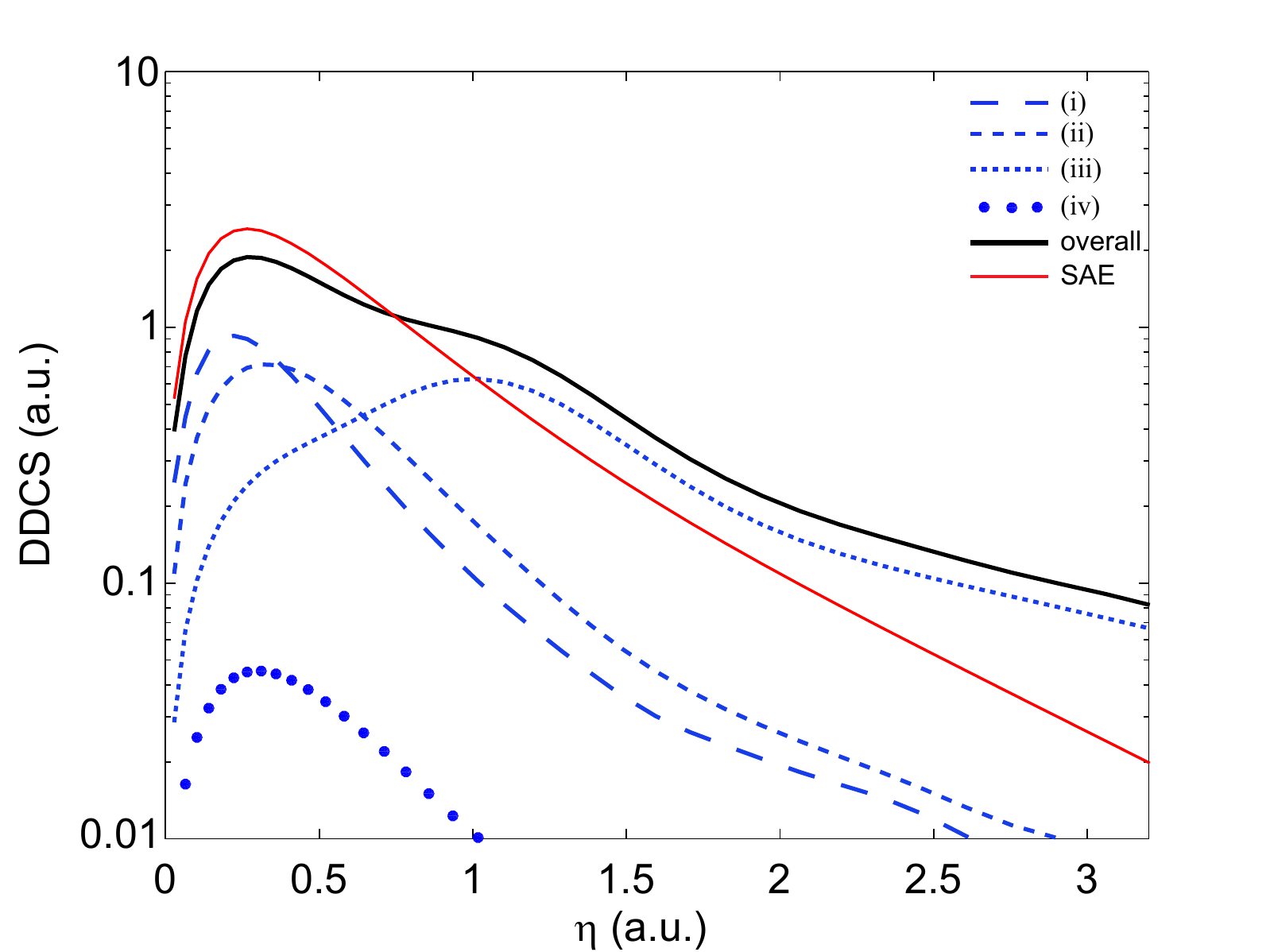}
\caption{(Color online) Same as Fig.~\ref{fig2} but for outgoing-electron energy $E_e= 10$~eV.}
\label{fig3}
\end{figure}

To elucidate the role of the multielectron processes for this collision system we now turn to a discussion of each process individually. 

\subsection{Contributions}
\begin{figure}[t]
\centering
\includegraphics[width=0.51\textwidth]{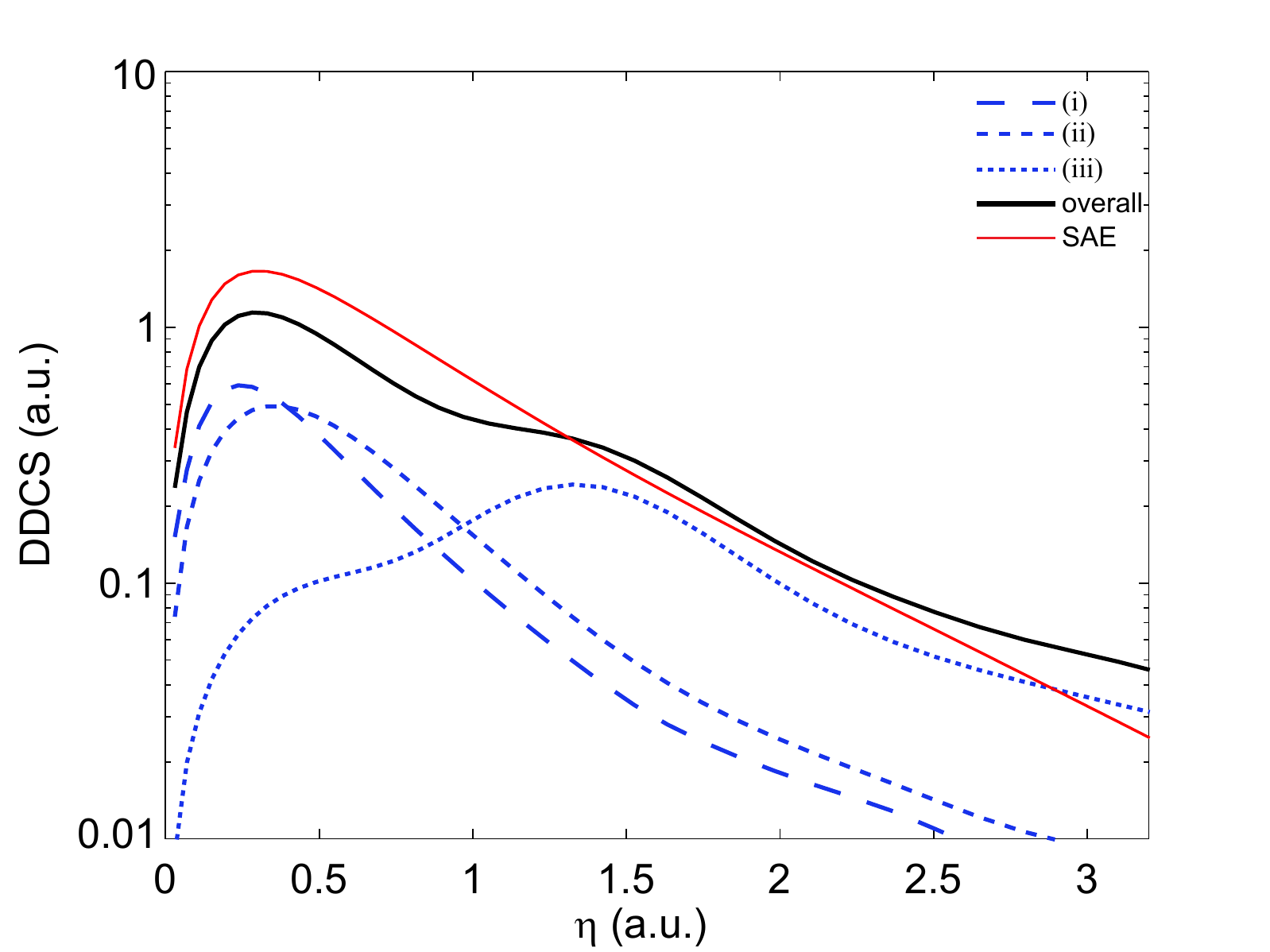}
\caption{(Color online) Same as Fig.~\ref{fig2} but for outgoing-electron energy $E_e= 20$~eV. The shake-off process has been removed due the negligible
 influence on the overall DDCS.} 
\label{fig4}
\end{figure}

Each process discussed in Sec.~\ref{Sec:theory} has a different dependence on the transverse momentum transfer $\eta$ for different outgoing-electron energy $E_e$. 
This results in a distinct variation of the magnitude of the DDCS for different $E_e$ and in different shapes of each DDCS as seen in Figs.~\ref{fig1a} and \ref{fig1b}. 
The contributions are depicted in Figs.~\ref{fig2},~\ref{fig3}, and~\ref{fig4} for $E_e=2$, 10, and 20~eV, respectively. 

The DDCSs for processes with the inner-shell electron detachment as the ionization channel, as well as the shake process [(i),(ii), and (iv)] resemble the SAE results, 
i.e., they have narrow maxima at low values of the momentum transfer and decrease rapidly with increasing $\eta$. The EI2 DDCS, on the other hand, peaks at larger $\eta$ and decreases less rapidly as $\eta$ increases. 
This behavior is inherited from the outer-shell ionization [see Figs. 1(c)-3(c) of Ref.~\cite{GulyasPRA2014} for the valence-shell ionization DDCSs] 
by combining the outer-shell ionization amplitude with amplitudes for inner-shell excitations and elastic scattering. 
While all the processes from (i) to (iv) play an important role in increasing the overall DDCS to match the experimental values at low $\eta$, it is the EI2 process that is responsible for the broadening of the DDCS
such that it agrees well with the experiment at $E_e=2$~eV (Fig.~\ref{fig2}). 

As the energy of the outgoing electron $E_e$ increases, the partial DDCS shapes for all the aforementioned processes, except the EI2 process, remain practically unchanged and only the overall strength diminishes.
In particular the shake off process becomes negligible at $E_e=20$~eV.
The DDCS for the EI2 process, on the other hand, is clearly suppressed, changing the shape most noticeably (see Figs.~\ref{fig2},~\ref{fig3}, and~\ref{fig4}): the maximum moves to higher momentum transfers 
as $E_e$ increases. 
This trend allows us to explain the broadening of the experimental DDCS, as well as the appearance of knees described in the previous subsection. 
Hence, it can be concluded that the shift of the maximum in the EI2 DDCS with increasing outgoing-electron energy is responsible for the broadening of the overall DDCS. This feature improves 
the agreement with the experimental data for most of the transverse momentum transfer range. 

Finally, we note that calculations for the H$^+$-Li system at 6~MeV/amu have shown that for weak perturbations the role of multielectron processes becomes negligible. A similar observation has been made for the Li$^{2+}$-Li system at 2.29~MeV/amu~\cite{SpiewanowskiJPCF2015}. 
In the latter case multielectron processes amounted to less than 10\% of the overall SDCS whereas for the H$^+$-Li system at 6~MeV/amu
these contributions are less than 1\%. The cause of the observed discrepancy between theory and experiment for these systems remains unknown, but can perhaps be attributed to 
electron correlations that are not accounted for in our model.

\section{Conclusions}\label{Sec:conclusions}

We presented a method to analyze multielectron processes by combining TC-BGM results for excitation with CDW-EIS results for ionization at the level of amplitudes. 
This approach allowed us to explain the experimental DDCSs for the O$^{8+}$-Li collision system at 1.5~MeV/amu. We attribute the leading role in broadening the DDCSs as compared to SAE CDW-EIS results to excitation-ionization processes, in particular, the EI2 process. 

Multielectron processes become less important as the perturbation parameter ($\beta=Z_p/v$) decreases. 
Hence, it would be interesting to see experimental results for systems with perturbation 
parameter between $\beta=1.03$ for the O$^{8+}$-Li collision system at 1.5~MeV/amu~\cite{LaForgeJPB2013} and $\beta=0.21$ for Li$^{2+}$-Li collisions at 2.29~MeV/amu~\cite{SpiewanowskiJPCF2015}.
This would allow us to trace how the role of the multielectron processes scales with the perturbation parameter.

\section*{Acknowledgements}
This work was supported by the Natural Sciences and Engineering Research Council of Canada (NSERC) under Grant No. RGPIN-2014-03611 and by the Hungarian Scientific Research Fund (OTKA Grant No. K 109440). 
We thank Dr. D. Fischer for providing details about the experimental data.

\renewcommand{\theequation}{A\arabic{equation}}
\setcounter{equation}{0} 
\section*{Appendix}
We show the validity of Eq.~\eqref{Pvacke2} by exploiting the approach given by Eq.~\eqref{A}.
The three-electron amplitude for a given set of spin projections is written as
\begin{equation}
 A_{\alpha_{1\uparrow}\alpha_{2\downarrow}\alpha_{3\uparrow}}=
 \begin{vmatrix}
 \langle \alpha_{1\uparrow}|\psi_{1s\uparrow} \rangle & 0 & \langle \alpha_{1\uparrow}|\psi_{2s\uparrow} \rangle \\
 0 & \langle \alpha_{2\downarrow}|\psi_{1s\downarrow} \rangle & 0 \\
 \langle \alpha_{3\uparrow}|\psi_{1s\uparrow} \rangle & 0 & \langle \alpha_{3\uparrow}|\psi_{2s\uparrow} \rangle \\
 \end{vmatrix},
\end{equation}
where we made use of the fact that due to the lack of spin dependence of the Hamiltonian projections of the final single-electron wave functions onto final states with an opposite spin projection are zero 
and, henceforth, we omit the spin projections for brevity with the exception of the subscripts in the amplitudes that follow to specify the final-state configuration.

To sketch the steps to get from Eq.~\eqref{Pvacke1} to Eq.~\eqref{Pvacke2} we write the amplitudes of interest as
\begin{eqnarray}\label{A1}
 A_{1s\uparrow f\downarrow \vec{k_e}\uparrow}&=&
 \begin{vmatrix}
\langle 1s|\psi_{1s} \rangle            &                         0                          & \langle 1s|\psi_{2s} \rangle \\
0                                                        & \langle f|\psi_{1s} \rangle &                 0 \\
\langle {\vec{k_e}}|\psi_{1s} \rangle &                         0                          & \langle {\vec{k_e}}|\psi_{2s} \rangle \\
\end{vmatrix}\nonumber\\
&=& \langle 1s|\psi_{1s} \rangle \langle f|\psi_{1s} \rangle \langle {\vec{k_e}}|\psi_{2s} \rangle\\
&&-\langle {\vec{k_e}}|\psi_{1s} \rangle \langle f|\psi_{1s} \rangle \langle 1s|\psi_{2s} \rangle\nonumber,
\end{eqnarray}

\begin{eqnarray}\label{A2}
 A_{f\uparrow 1s\downarrow  \vec{k_e}\uparrow}&=&
  \begin{vmatrix}
\langle f|\psi_{1s} \rangle            &                         0                          & \langle f|\psi_{2s} \rangle \\
0                                                        & \langle 1s|\psi_{1s} \rangle &                 0 \\
\langle {\vec{k_e}}|\psi_{1s} \rangle &                         0                          & \langle {\vec{k_e}}|\psi_{2s} \rangle \\
\end{vmatrix}\nonumber\\
&=&  \langle f|\psi_{1s} \rangle \langle 1s|\psi_{1s} \rangle \langle {\vec{k_e}}|\psi_{2s} \rangle\\
&&-\langle {\vec{k_e}}|\psi_{1s} \rangle \langle 1s|\psi_{1s} \rangle \langle f|\psi_{2s} \rangle\nonumber,
\end{eqnarray}

\begin{eqnarray}\label{A3}
 A_{1s\uparrow\vec{k_e}\downarrow f\uparrow}&=&
  \begin{vmatrix}
\langle 1s|\psi_{1s} \rangle            &                         0                          & \langle 1s|\psi_{2s} \rangle \\
0                                                        & \langle {\vec{k_e}}|\psi_{1s} \rangle &                 0 \\
\langle f|\psi_{1s} \rangle &                         0                          & \langle f|\psi_{2s} \rangle \\
\end{vmatrix}\nonumber\\
&=& \langle 1s|\psi_{1s} \rangle \langle {\vec{k_e}}|\psi_{1s} \rangle \langle f|\psi_{2s} \rangle\\
&&-\langle f|\psi_{1s} \rangle \langle {\vec{k_e}}|\psi_{1s} \rangle \langle 1s|\psi_{2s} \rangle\nonumber.
\end{eqnarray}
Using Eqs.~\eqref{A1}-\eqref{A3}, we can express Eq.~\eqref{Pvacke1} via Eqs.~\eqref{DDCS}-\eqref{aik} as
\begin{widetext}
\begin{eqnarray}
 \frac{\mathrm{d}\sigma^\mathrm{vac}}{\mathrm{d}E_e\mathrm{d}\eta}&=& 
  \sum\limits_{f\neq 1s}\left(\frac{\mathrm{d}\sigma{\substack{1s\rightarrow1s\\ 1s\rightarrow f\\ 2s\rightarrow \vec{k_e}}}}{\mathrm{d}E_e\mathrm{d}\eta}
 +\frac{\mathrm{d}\sigma{\substack{1s\rightarrow \vec{k_e}\\ 1s\rightarrow f\\ 2s\rightarrow 1s}}}{\mathrm{d}E_e\mathrm{d}\eta}
 +\frac{\mathrm{d}\sigma{\substack{1s\rightarrow f\\ 1s\rightarrow 1s\\ 2s\rightarrow \vec{k_e}}}}{\mathrm{d}E_e\mathrm{d}\eta}
 +\frac{\mathrm{d}\sigma{\substack{1s\rightarrow 1s\\ 1s\rightarrow \vec{k_e}\\ 2s\rightarrow f}}}{\mathrm{d}E_e\mathrm{d}\eta}
 +\frac{\mathrm{d}\sigma{\substack{1s\rightarrow \vec{k_e}\\ 1s\rightarrow 1s\\ 2s\rightarrow f}}}{\mathrm{d}E_e\mathrm{d}\eta} 
 +\frac{\mathrm{d}\sigma{\substack{1s\rightarrow f\\ 1s\rightarrow \vec{k_e}\\ 2s\rightarrow 1s}}}{\mathrm{d}E_e\mathrm{d}\eta}   
 +\Delta^f_{E_e,\eta}\right)\\
 &=&\sum\limits_{f\neq 1s}\left(2\frac{\mathrm{d}\sigma{\substack{1s\rightarrow1s\\ 1s\rightarrow f\\ 2s\rightarrow \vec{k_e}}}}{\mathrm{d}E_e\mathrm{d}\eta}
  +2\frac{\mathrm{d}\sigma{\substack{1s\rightarrow \vec{k_e}\\ 1s\rightarrow f\\ 2s\rightarrow 1s}}}{\mathrm{d}E_e\mathrm{d}\eta}
  +2\frac{\mathrm{d}\sigma{\substack{1s\rightarrow1s\\ 1s\rightarrow \vec{k_e}\\ 2s\rightarrow f}}}{\mathrm{d}E_e\mathrm{d}\eta}+\Delta^f_{E_e,\eta}\right)\label{fac2}\\
 &=&\frac{\mathrm{d}\sigma^\mathrm{excl}}{\mathrm{d}E_e\mathrm{d}\eta}+\sum\limits_{\mathclap{f\neq 1s,2s}}\frac{\mathrm{d}\sigma^\mathrm{EI1}_{f}}{\mathrm{d}E_e\mathrm{d}\eta}
    +\sum\limits_{f\neq 1s}\left(\frac{\mathrm{d}\sigma^\mathrm{EI2}_{f}}{\mathrm{d}E_e\mathrm{d}\eta}\
    +\frac{\mathrm{d}\sigma^\mathrm{ex}_{f}}{\mathrm{d}E_e\mathrm{d}\eta} + \Delta^f_{E_e,\eta}\right),
 \end{eqnarray}
\end{widetext}
where $\frac{\mathrm{d}\sigma{\substack{1s\rightarrow \alpha_1\\ 1s\rightarrow \alpha_2\\ 2s\rightarrow \alpha_3}}}{\mathrm{d}E_e\mathrm{d}\eta}$ are the DDCSs 
for the transitions to the $|\alpha_1\alpha_2\alpha_3\rangle$ final state and $\Delta^f_{E_e,\eta}$ consists of all the cross terms. 
Finally, the factors of two in Eq.~\eqref{fac2} are incorporated into the respective amplitudes, resulting in $\sqrt{2}$ factors in Eqs.~\eqref{excl}-\eqref{EI2}.


\begin{thebibliography}{18}%
\makeatletter
\providecommand \@ifxundefined [1]{%
 \@ifx{#1\undefined}
}%
\providecommand \@ifnum [1]{%
 \ifnum #1\expandafter \@firstoftwo
 \else \expandafter \@secondoftwo
 \fi
}%
\providecommand \@ifx [1]{%
 \ifx #1\expandafter \@firstoftwo
 \else \expandafter \@secondoftwo
 \fi
}%
\providecommand \natexlab [1]{#1}%
\providecommand \enquote  [1]{``#1''}%
\providecommand \bibnamefont  [1]{#1}%
\providecommand \bibfnamefont [1]{#1}%
\providecommand \citenamefont [1]{#1}%
\providecommand \href@noop [0]{\@secondoftwo}%
\providecommand \href [0]{\begingroup \@sanitize@url \@href}%
\providecommand \@href[1]{\@@startlink{#1}\@@href}%
\providecommand \@@href[1]{\endgroup#1\@@endlink}%
\providecommand \@sanitize@url [0]{\catcode `\\12\catcode `\$12\catcode
  `\&12\catcode `\#12\catcode `\^12\catcode `\_12\catcode `\%12\relax}%
\providecommand \@@startlink[1]{}%
\providecommand \@@endlink[0]{}%
\providecommand \url  [0]{\begingroup\@sanitize@url \@url }%
\providecommand \@url [1]{\endgroup\@href {#1}{\urlprefix }}%
\providecommand \urlprefix  [0]{URL }%
\providecommand \Eprint [0]{\href }%
\providecommand \doibase [0]{http://dx.doi.org/}%
\providecommand \selectlanguage [0]{\@gobble}%
\providecommand \bibinfo  [0]{\@secondoftwo}%
\providecommand \bibfield  [0]{\@secondoftwo}%
\providecommand \translation [1]{[#1]}%
\providecommand \BibitemOpen [0]{}%
\providecommand \bibitemStop [0]{}%
\providecommand \bibitemNoStop [0]{.\EOS\space}%
\providecommand \EOS [0]{\spacefactor3000\relax}%
\providecommand \BibitemShut  [1]{\csname bibitem#1\endcsname}%
\let\auto@bib@innerbib\@empty
\bibitem [{\citenamefont {Schulz}\ and\ \citenamefont
  {Madison}(2006)}]{Schulz2006}%
  \BibitemOpen
  \bibfield  {author} {\bibinfo {author} {\bibfnamefont {M.}~\bibnamefont
  {Schulz}}\ and\ \bibinfo {author} {\bibfnamefont {D.~H.}\ \bibnamefont
  {Madison}},\ }\bibfield  {title} {\enquote {\bibinfo {title} {Studies of the
  few-body problem in atomic break-up processes},}\ }\href@noop {} {\bibfield
  {journal} {\bibinfo  {journal} {Int. J. Mod. Phys. A}\ }\textbf {\bibinfo
  {volume} {21}},\ \bibinfo {pages} {3649--3672} (\bibinfo {year}
  {2006})}\BibitemShut {NoStop}%
\bibitem [{\citenamefont {Ullrich}\ \emph {et~al.}(2003)\citenamefont
  {Ullrich}, \citenamefont {Moshammer}, \citenamefont {Dorn}, \citenamefont
  {D\"orner}, \citenamefont {Schmidt},\ and\ \citenamefont
  {Schmidt-B\"ocking}}]{UllrichRepProgPhys2003}%
  \BibitemOpen
  \bibfield  {author} {\bibinfo {author} {\bibfnamefont {J.}~\bibnamefont
  {Ullrich}}, \bibinfo {author} {\bibfnamefont {R.}~\bibnamefont {Moshammer}},
  \bibinfo {author} {\bibfnamefont {A.}~\bibnamefont {Dorn}}, \bibinfo {author}
  {\bibfnamefont {R.}~\bibnamefont {D\"orner}}, \bibinfo {author}
  {\bibfnamefont {L.~Ph.~H.}\ \bibnamefont {Schmidt}}, \ and\ \bibinfo {author}
  {\bibfnamefont {H.}~\bibnamefont {Schmidt-B\"ocking}},\ }\bibfield  {title}
  {\enquote {\bibinfo {title} {Recoil-ion and electron momentum spectroscopy:
  reaction-microscopes},}\ }\href
  {http://stacks.iop.org/0034-4885/66/i=9/a=203} {\bibfield  {journal}
  {\bibinfo  {journal} {Rep. Prog. Phys.}\ }\textbf {\bibinfo {volume} {66}},\
  \bibinfo {pages} {1463} (\bibinfo {year} {2003})}\BibitemShut {NoStop}%
\bibitem [{\citenamefont {Schulz}\ \emph {et~al.}(2003)\citenamefont {Schulz},
  \citenamefont {Moshammer}, \citenamefont {Fischer}, \citenamefont {Kollmus},
  \citenamefont {Madison}, \citenamefont {Jones},\ and\ \citenamefont
  {Ullrich}}]{SchulzNature2003}%
  \BibitemOpen
  \bibfield  {author} {\bibinfo {author} {\bibfnamefont {M.}~\bibnamefont
  {Schulz}}, \bibinfo {author} {\bibfnamefont {R.}~\bibnamefont {Moshammer}},
  \bibinfo {author} {\bibfnamefont {D.}~\bibnamefont {Fischer}}, \bibinfo
  {author} {\bibfnamefont {H.}~\bibnamefont {Kollmus}}, \bibinfo {author}
  {\bibfnamefont {D.~H.}\ \bibnamefont {Madison}}, \bibinfo {author}
  {\bibfnamefont {S.}~\bibnamefont {Jones}}, \ and\ \bibinfo {author}
  {\bibfnamefont {J.}~\bibnamefont {Ullrich}},\ }\bibfield  {title} {\enquote
  {\bibinfo {title} {Three-dimensional imaging of atomic four-body
  processes},}\ }\href@noop {} {\bibfield  {journal} {\bibinfo  {journal}
  {Nature}\ }\textbf {\bibinfo {volume} {422}},\ \bibinfo {pages} {48--50}
  (\bibinfo {year} {2003})}\BibitemShut {NoStop}%
\bibitem [{\citenamefont {Fischer}\ \emph {et~al.}(2012)\citenamefont
  {Fischer}, \citenamefont {Globig}, \citenamefont {Goullon}, \citenamefont
  {Grieser}, \citenamefont {Hubele}, \citenamefont {de~Jesus}, \citenamefont
  {Kelkar}, \citenamefont {LaForge}, \citenamefont {Lindenblatt}, \citenamefont
  {Misra}, \citenamefont {Najjari}, \citenamefont {Schneider}, \citenamefont
  {Schulz}, \citenamefont {Sell},\ and\ \citenamefont {Wang}}]{FischerPRL2012}%
  \BibitemOpen
  \bibfield  {author} {\bibinfo {author} {\bibfnamefont {D.}~\bibnamefont
  {Fischer}}, \bibinfo {author} {\bibfnamefont {D.}~\bibnamefont {Globig}},
  \bibinfo {author} {\bibfnamefont {J.}~\bibnamefont {Goullon}}, \bibinfo
  {author} {\bibfnamefont {M.}~\bibnamefont {Grieser}}, \bibinfo {author}
  {\bibfnamefont {R.}~\bibnamefont {Hubele}}, \bibinfo {author} {\bibfnamefont
  {V.~L.~B.}\ \bibnamefont {de~Jesus}}, \bibinfo {author} {\bibfnamefont
  {A.}~\bibnamefont {Kelkar}}, \bibinfo {author} {\bibfnamefont
  {A.}~\bibnamefont {LaForge}}, \bibinfo {author} {\bibfnamefont
  {H.}~\bibnamefont {Lindenblatt}}, \bibinfo {author} {\bibfnamefont
  {D.}~\bibnamefont {Misra}}, \bibinfo {author} {\bibfnamefont
  {B.}~\bibnamefont {Najjari}}, \bibinfo {author} {\bibfnamefont
  {K.}~\bibnamefont {Schneider}}, \bibinfo {author} {\bibfnamefont
  {M.}~\bibnamefont {Schulz}}, \bibinfo {author} {\bibfnamefont
  {M.}~\bibnamefont {Sell}}, \ and\ \bibinfo {author} {\bibfnamefont
  {X.}~\bibnamefont {Wang}},\ }\bibfield  {title} {\enquote {\bibinfo {title}
  {Ion-lithium collision dynamics studied with a laser-cooled in-ring
  target},}\ }\href {\doibase 10.1103/PhysRevLett.109.113202} {\bibfield
  {journal} {\bibinfo  {journal} {Phys. Rev. Lett.}\ }\textbf {\bibinfo
  {volume} {109}},\ \bibinfo {pages} {113202} (\bibinfo {year}
  {2012})}\BibitemShut {NoStop}%
\bibitem [{\citenamefont {Hubele}\ \emph {et~al.}(2015)\citenamefont {Hubele},
  \citenamefont {Schuricke}, \citenamefont {Goullon}, \citenamefont
  {Lindenblatt}, \citenamefont {Ferreira}, \citenamefont {Laforge},
  \citenamefont {Brühl}, \citenamefont {de~Jesus}, \citenamefont {Globig},
  \citenamefont {Kelkar}, \citenamefont {Misra}, \citenamefont {Schneider},
  \citenamefont {Schulz}, \citenamefont {Sell}, \citenamefont {Song},
  \citenamefont {Wang}, \citenamefont {Zhang},\ and\ \citenamefont
  {Fischer}}]{HubeleRSI2015}%
  \BibitemOpen
  \bibfield  {author} {\bibinfo {author} {\bibfnamefont {R.}~\bibnamefont
  {Hubele}}, \bibinfo {author} {\bibfnamefont {M.}~\bibnamefont {Schuricke}},
  \bibinfo {author} {\bibfnamefont {J.}~\bibnamefont {Goullon}}, \bibinfo
  {author} {\bibfnamefont {H.}~\bibnamefont {Lindenblatt}}, \bibinfo {author}
  {\bibfnamefont {N.}~\bibnamefont {Ferreira}}, \bibinfo {author}
  {\bibfnamefont {A.}~\bibnamefont {Laforge}}, \bibinfo {author} {\bibfnamefont
  {E.}~\bibnamefont {Brühl}}, \bibinfo {author} {\bibfnamefont {V.~L.~B.}\
  \bibnamefont {de~Jesus}}, \bibinfo {author} {\bibfnamefont {D.}~\bibnamefont
  {Globig}}, \bibinfo {author} {\bibfnamefont {A.}~\bibnamefont {Kelkar}},
  \bibinfo {author} {\bibfnamefont {D.}~\bibnamefont {Misra}}, \bibinfo
  {author} {\bibfnamefont {K.}~\bibnamefont {Schneider}}, \bibinfo {author}
  {\bibfnamefont {M.}~\bibnamefont {Schulz}}, \bibinfo {author} {\bibfnamefont
  {M.}~\bibnamefont {Sell}}, \bibinfo {author} {\bibfnamefont {Z.}~\bibnamefont
  {Song}}, \bibinfo {author} {\bibfnamefont {X.}~\bibnamefont {Wang}}, \bibinfo
  {author} {\bibfnamefont {S.}~\bibnamefont {Zhang}}, \ and\ \bibinfo {author}
  {\bibfnamefont {D.}~\bibnamefont {Fischer}},\ }\bibfield  {title} {\enquote
  {\bibinfo {title} {Electron and recoil ion momentum imaging with a
  magneto-optically trapped target},}\ }\href {\doibase
  http://dx.doi.org/10.1063/1.4914040} {\bibfield  {journal} {\bibinfo
  {journal} {Rev. Sci. Instrum.}\ }\textbf {\bibinfo {volume} {86}},\ \bibinfo
  {eid} {033105} (\bibinfo {year} {2015})}\BibitemShut {NoStop}%
\bibitem [{\citenamefont {LaForge}\ \emph {et~al.}(2013)\citenamefont
  {LaForge}, \citenamefont {Hubele}, \citenamefont {Goullon}, \citenamefont
  {Wang}, \citenamefont {Schneider}, \citenamefont {de~Jesus}, \citenamefont
  {Najjari}, \citenamefont {Voitkiv}, \citenamefont {Grieser}, \citenamefont
  {Schulz},\ and\ \citenamefont {Fischer}}]{LaForgeJPB2013}%
  \BibitemOpen
  \bibfield  {author} {\bibinfo {author} {\bibfnamefont {A.~C.}\ \bibnamefont
  {LaForge}}, \bibinfo {author} {\bibfnamefont {R.}~\bibnamefont {Hubele}},
  \bibinfo {author} {\bibfnamefont {J.}~\bibnamefont {Goullon}}, \bibinfo
  {author} {\bibfnamefont {X.}~\bibnamefont {Wang}}, \bibinfo {author}
  {\bibfnamefont {K.}~\bibnamefont {Schneider}}, \bibinfo {author}
  {\bibfnamefont {V.~L.~B.}\ \bibnamefont {de~Jesus}}, \bibinfo {author}
  {\bibfnamefont {B.}~\bibnamefont {Najjari}}, \bibinfo {author} {\bibfnamefont
  {A.~B.}\ \bibnamefont {Voitkiv}}, \bibinfo {author} {\bibfnamefont
  {M.}~\bibnamefont {Grieser}}, \bibinfo {author} {\bibfnamefont
  {M.}~\bibnamefont {Schulz}}, \ and\ \bibinfo {author} {\bibfnamefont
  {D.}~\bibnamefont {Fischer}},\ }\bibfield  {title} {\enquote {\bibinfo
  {title} {Initial-state selective study of ionization dynamics in ion-{L}i
  collisions},}\ }\href {http://stacks.iop.org/0953-4075/46/i=3/a=031001}
  {\bibfield  {journal} {\bibinfo  {journal} {J. Phys. B}\ }\textbf {\bibinfo
  {volume} {46}},\ \bibinfo {pages} {031001} (\bibinfo {year}
  {2013})}\BibitemShut {NoStop}%
\bibitem [{\citenamefont {Hubele}\ \emph {et~al.}(2013)\citenamefont {Hubele},
  \citenamefont {LaForge}, \citenamefont {Schulz}, \citenamefont {Goullon},
  \citenamefont {Wang}, \citenamefont {Najjari}, \citenamefont {Ferreira},
  \citenamefont {Grieser}, \citenamefont {de~Jesus}, \citenamefont {Moshammer},
  \citenamefont {Schneider}, \citenamefont {Voitkiv},\ and\ \citenamefont
  {Fischer}}]{HubelePRL2013}%
  \BibitemOpen
  \bibfield  {author} {\bibinfo {author} {\bibfnamefont {R.}~\bibnamefont
  {Hubele}}, \bibinfo {author} {\bibfnamefont {A.}~\bibnamefont {LaForge}},
  \bibinfo {author} {\bibfnamefont {M.}~\bibnamefont {Schulz}}, \bibinfo
  {author} {\bibfnamefont {J.}~\bibnamefont {Goullon}}, \bibinfo {author}
  {\bibfnamefont {X.}~\bibnamefont {Wang}}, \bibinfo {author} {\bibfnamefont
  {B.}~\bibnamefont {Najjari}}, \bibinfo {author} {\bibfnamefont
  {N.}~\bibnamefont {Ferreira}}, \bibinfo {author} {\bibfnamefont
  {M.}~\bibnamefont {Grieser}}, \bibinfo {author} {\bibfnamefont {V.~L.~B.}\
  \bibnamefont {de~Jesus}}, \bibinfo {author} {\bibfnamefont {R.}~\bibnamefont
  {Moshammer}}, \bibinfo {author} {\bibfnamefont {K.}~\bibnamefont
  {Schneider}}, \bibinfo {author} {\bibfnamefont {A.~B.}\ \bibnamefont
  {Voitkiv}}, \ and\ \bibinfo {author} {\bibfnamefont {D.}~\bibnamefont
  {Fischer}},\ }\bibfield  {title} {\enquote {\bibinfo {title} {Polarization
  and interference effects in ionization of li by ion impact},}\ }\href
  {\doibase 10.1103/PhysRevLett.110.133201} {\bibfield  {journal} {\bibinfo
  {journal} {Phys. Rev. Lett.}\ }\textbf {\bibinfo {volume} {110}},\ \bibinfo
  {pages} {133201} (\bibinfo {year} {2013})}\BibitemShut {NoStop}%
\bibitem [{\citenamefont {\'Spiewanowski}\ \emph {et~al.}(2015)\citenamefont
  {\'Spiewanowski}, \citenamefont {Guly\'as}, \citenamefont {Horbatsch},
  \citenamefont {Goullon}, \citenamefont {Ferreira}, \citenamefont {Hubele},
  \citenamefont {de~Jesus}, \citenamefont {Lindenblatt}, \citenamefont
  {Schneider}, \citenamefont {Schulz}, \citenamefont {Schuricke}, \citenamefont
  {Song}, \citenamefont {Zhang}, \citenamefont {Fischer},\ and\ \citenamefont
  {Kirchner}}]{SpiewanowskiJPCF2015}%
  \BibitemOpen
  \bibfield  {author} {\bibinfo {author} {\bibfnamefont {M.~D.}\ \bibnamefont
  {\'Spiewanowski}}, \bibinfo {author} {\bibfnamefont {L.}~\bibnamefont
  {Guly\'as}}, \bibinfo {author} {\bibfnamefont {M.}~\bibnamefont {Horbatsch}},
  \bibinfo {author} {\bibfnamefont {J.}~\bibnamefont {Goullon}}, \bibinfo
  {author} {\bibfnamefont {N.}~\bibnamefont {Ferreira}}, \bibinfo {author}
  {\bibfnamefont {R.}~\bibnamefont {Hubele}}, \bibinfo {author} {\bibfnamefont
  {V.~L.~B.}\ \bibnamefont {de~Jesus}}, \bibinfo {author} {\bibfnamefont
  {H.}~\bibnamefont {Lindenblatt}}, \bibinfo {author} {\bibfnamefont
  {K.}~\bibnamefont {Schneider}}, \bibinfo {author} {\bibfnamefont
  {M.}~\bibnamefont {Schulz}}, \bibinfo {author} {\bibfnamefont
  {M.}~\bibnamefont {Schuricke}}, \bibinfo {author} {\bibfnamefont
  {Z.}~\bibnamefont {Song}}, \bibinfo {author} {\bibfnamefont {S.}~\bibnamefont
  {Zhang}}, \bibinfo {author} {\bibfnamefont {D.}~\bibnamefont {Fischer}}, \
  and\ \bibinfo {author} {\bibfnamefont {T.}~\bibnamefont {Kirchner}},\
  }\bibfield  {title} {\enquote {\bibinfo {title} {Target electron ionization
  in {L}i$^{2+}$-{L}i collisions: A multi-electron perspective},}\ }\href
  {http://stacks.iop.org/1742-6596/601/i=1/a=012010} {\bibfield  {journal}
  {\bibinfo  {journal} {J. Phys. Conf. Series}\ }\textbf {\bibinfo {volume}
  {601}},\ \bibinfo {pages} {012010} (\bibinfo {year} {2015})}\BibitemShut
  {NoStop}%
\bibitem [{\citenamefont {Ciappina}\ \emph {et~al.}(2013)\citenamefont
  {Ciappina}, \citenamefont {Pindzola},\ and\ \citenamefont
  {Colgan}}]{CiappinaPRA2013}%
  \BibitemOpen
  \bibfield  {author} {\bibinfo {author} {\bibfnamefont {M.~F.}\ \bibnamefont
  {Ciappina}}, \bibinfo {author} {\bibfnamefont {M.~S.}\ \bibnamefont
  {Pindzola}}, \ and\ \bibinfo {author} {\bibfnamefont {J.}~\bibnamefont
  {Colgan}},\ }\bibfield  {title} {\enquote {\bibinfo {title} {Fully
  differential cross section for {O}${}^{8+}$-impact ionization of {L}i},}\
  }\href {\doibase 10.1103/PhysRevA.87.042706} {\bibfield  {journal} {\bibinfo
  {journal} {Phys. Rev. A}\ }\textbf {\bibinfo {volume} {87}},\ \bibinfo
  {pages} {042706} (\bibinfo {year} {2013})}\BibitemShut {NoStop}%
\bibitem [{\citenamefont {Walters}\ and\ \citenamefont
  {Whelan}(2014)}]{WaltersPRA2014}%
  \BibitemOpen
  \bibfield  {author} {\bibinfo {author} {\bibfnamefont {H.~R.~J.}\
  \bibnamefont {Walters}}\ and\ \bibinfo {author} {\bibfnamefont {Colm~T.}\
  \bibnamefont {Whelan}},\ }\bibfield  {title} {\enquote {\bibinfo {title}
  {Differential ionization of {L}i(2$s$) and {L}i(2$p$) under proton and
  {O}${}^{8+}$ impact},}\ }\href {\doibase 10.1103/PhysRevA.89.032709}
  {\bibfield  {journal} {\bibinfo  {journal} {Phys. Rev. A}\ }\textbf {\bibinfo
  {volume} {89}},\ \bibinfo {pages} {032709} (\bibinfo {year}
  {2014})}\BibitemShut {NoStop}%
\bibitem [{\citenamefont {Kirchner}\ \emph {et~al.}(2014)\citenamefont
  {Kirchner}, \citenamefont {Khazai},\ and\ \citenamefont
  {Guly\'as}}]{KirchnerPRA2014}%
  \BibitemOpen
  \bibfield  {author} {\bibinfo {author} {\bibfnamefont {T.}~\bibnamefont
  {Kirchner}}, \bibinfo {author} {\bibfnamefont {N.}~\bibnamefont {Khazai}}, \
  and\ \bibinfo {author} {\bibfnamefont {L.}~\bibnamefont {Guly\'as}},\
  }\bibfield  {title} {\enquote {\bibinfo {title} {Role of two-electron
  excitation-ionization processes in the ionization of lithium atoms by fast
  ion impact},}\ }\href {\doibase 10.1103/PhysRevA.89.062702} {\bibfield
  {journal} {\bibinfo  {journal} {Phys. Rev. A}\ }\textbf {\bibinfo {volume}
  {89}},\ \bibinfo {pages} {062702} (\bibinfo {year} {2014})}\BibitemShut
  {NoStop}%
\bibitem [{\citenamefont {Guly\'as}\ \emph {et~al.}(2014)\citenamefont
  {Guly\'as}, \citenamefont {Egri},\ and\ \citenamefont
  {Kirchner}}]{GulyasPRA2014}%
  \BibitemOpen
  \bibfield  {author} {\bibinfo {author} {\bibfnamefont {L.}~\bibnamefont
  {Guly\'as}}, \bibinfo {author} {\bibfnamefont {S.}~\bibnamefont {Egri}}, \
  and\ \bibinfo {author} {\bibfnamefont {T.}~\bibnamefont {Kirchner}},\
  }\bibfield  {title} {\enquote {\bibinfo {title} {Differential cross sections
  for single ionization of {L}i in collisions with fast protons and {O}$^{8+}$
  ions},}\ }\href {\doibase 10.1103/PhysRevA.90.062710} {\bibfield  {journal}
  {\bibinfo  {journal} {Phys. Rev. A}\ }\textbf {\bibinfo {volume} {90}},\
  \bibinfo {pages} {062710} (\bibinfo {year} {2014})}\BibitemShut {NoStop}%
\bibitem [{\citenamefont {Zapukhlyak}\ \emph {et~al.}(2005)\citenamefont
  {Zapukhlyak}, \citenamefont {Kirchner}, \citenamefont {L\"udde},
  \citenamefont {Knoop}, \citenamefont {Morgenstern},\ and\ \citenamefont
  {Hoekstra}}]{ZapukhlyakJPB2005}%
  \BibitemOpen
  \bibfield  {author} {\bibinfo {author} {\bibfnamefont {M.}~\bibnamefont
  {Zapukhlyak}}, \bibinfo {author} {\bibfnamefont {T.}~\bibnamefont
  {Kirchner}}, \bibinfo {author} {\bibfnamefont {H.~J.}\ \bibnamefont
  {L\"udde}}, \bibinfo {author} {\bibfnamefont {S.}~\bibnamefont {Knoop}},
  \bibinfo {author} {\bibfnamefont {R.}~\bibnamefont {Morgenstern}}, \ and\
  \bibinfo {author} {\bibfnamefont {R.}~\bibnamefont {Hoekstra}},\ }\bibfield
  {title} {\enquote {\bibinfo {title} {Inner- and outer-shell electron dynamics
  in proton collisions with sodium atoms},}\ }\href@noop {} {\bibfield
  {journal} {\bibinfo  {journal} {J. Phys. B}\ }\textbf {\bibinfo {volume}
  {38}},\ \bibinfo {pages} {2353} (\bibinfo {year} {2005})}\BibitemShut
  {NoStop}%
\bibitem [{\citenamefont {Guly\'as}\ \emph {et~al.}(1995)\citenamefont
  {Guly\'as}, \citenamefont {Fainstein},\ and\ \citenamefont
  {Salin}}]{GulyasJPB1995}%
  \BibitemOpen
  \bibfield  {author} {\bibinfo {author} {\bibfnamefont {L.}~\bibnamefont
  {Guly\'as}}, \bibinfo {author} {\bibfnamefont {P.~D.}\ \bibnamefont
  {Fainstein}}, \ and\ \bibinfo {author} {\bibfnamefont {A.}~\bibnamefont
  {Salin}},\ }\bibfield  {title} {\enquote {\bibinfo {title} {{CDW}-{EIS}
  theory of ionization by ion impact with {H}artree-{F}ock description of the
  target},}\ }\href@noop {} {\bibfield  {journal} {\bibinfo  {journal} {J.
  Phys. B}\ }\textbf {\bibinfo {volume} {28}},\ \bibinfo {pages} {245}
  (\bibinfo {year} {1995})}\BibitemShut {NoStop}%
\bibitem [{\citenamefont {Zapukhlyak}\ \emph {et~al.}(2008)\citenamefont
  {Zapukhlyak}, \citenamefont {Kirchner}, \citenamefont {Hasan}, \citenamefont
  {Tooke},\ and\ \citenamefont {Schulz}}]{ZapukhlyakJPB2008}%
  \BibitemOpen
  \bibfield  {author} {\bibinfo {author} {\bibfnamefont {M.}~\bibnamefont
  {Zapukhlyak}}, \bibinfo {author} {\bibfnamefont {T.}~\bibnamefont
  {Kirchner}}, \bibinfo {author} {\bibfnamefont {A.}~\bibnamefont {Hasan}},
  \bibinfo {author} {\bibfnamefont {B.}~\bibnamefont {Tooke}}, \ and\ \bibinfo
  {author} {\bibfnamefont {M.}~\bibnamefont {Schulz}},\ }\bibfield  {title}
  {\enquote {\bibinfo {title} {Projectile angular-differential cross sections
  for transfer and transfer excitation in proton collisions with helium},}\
  }\href {\doibase 10.1103/PhysRevA.77.012720} {\bibfield  {journal} {\bibinfo
  {journal} {Phys. Rev. A}\ }\textbf {\bibinfo {volume} {77}},\ \bibinfo
  {pages} {012720} (\bibinfo {year} {2008})}\BibitemShut {NoStop}%
\bibitem [{\citenamefont {L\"udde}\ and\ \citenamefont
  {Dreizler}(1985)}]{LuddeJPB1985}%
  \BibitemOpen
  \bibfield  {author} {\bibinfo {author} {\bibfnamefont {H.~J.}\ \bibnamefont
  {L\"udde}}\ and\ \bibinfo {author} {\bibfnamefont {R.~M.}\ \bibnamefont
  {Dreizler}},\ }\bibfield  {title} {\enquote {\bibinfo {title} {Comment on
  inclusive cross sections},}\ }\href
  {http://stacks.iop.org/0022-3700/18/i=1/a=012} {\bibfield  {journal}
  {\bibinfo  {journal} {J. Phys. B}\ }\textbf {\bibinfo {volume} {18}},\
  \bibinfo {pages} {107} (\bibinfo {year} {1985})}\BibitemShut {NoStop}%
\bibitem [{\citenamefont {Engel}\ and\ \citenamefont
  {Vosko}(1993)}]{EngelPRA1993}%
  \BibitemOpen
  \bibfield  {author} {\bibinfo {author} {\bibfnamefont {E.}~\bibnamefont
  {Engel}}\ and\ \bibinfo {author} {\bibfnamefont {S.~H.}\ \bibnamefont
  {Vosko}},\ }\bibfield  {title} {\enquote {\bibinfo {title} {Accurate
  optimized-potential-model solutions for spherical spin-polarized atoms:
  Evidence for limitations of the exchange-only local spin-density and
  generalized-gradient approximations},}\ }\href {\doibase
  10.1103/PhysRevA.47.2800} {\bibfield  {journal} {\bibinfo  {journal} {Phys.
  Rev. A}\ }\textbf {\bibinfo {volume} {47}},\ \bibinfo {pages} {2800--2811}
  (\bibinfo {year} {1993})}\BibitemShut {NoStop}%
\bibitem [{pri()}]{privateFisher}%
  \BibitemOpen
  \href@noop {} {\ }\bibinfo {note} {{D}. {F}ischer (private
  communication)}\BibitemShut {NoStop}%
\end{thebibliography}
\end{document}